# Image Segmentation and Processing for Efficient Parking Space Analysis

Chetan Sai Tutika, Charan Vallapaneni, Karthik R, Bharath KP,
N Ruban Rajesh Kumar Muthu Senior *Member, IEEE*
School of Electronics Engineering
VIT University, India chetansai.tutika@gmail.com

Email: chetansai.tutika@gmail.com, charanvallapaneni@gmail.com, tkgravikarthik@gmail.com,
bharathkp25@gmail.com,nruban@vit.ac.in,mrajeshkumar@vit.ac.in

*Abstract*—In this paper, we develop a method to detect vacant parking spaces in an environment with unclear segments and contours with the help of MATLAB's image processing capabilities. Due to the anomalies present in the parking spaces,such as uneven illumination, distorted slot lines and overlapping of cars. The present day conventional algorithms have difficulties processing the image for accurate results. The algorithm proposed uses a combination of image pre-processing and false contour detection techniques to improve the detection efficiency. The proposed method also eliminates the need to employ individual sensors to detect a car, instead uses real-time static images to consider a group of slots together, instead of the usual single slot method. This greatly decreases the expenses required to design an efficient parking system. We compare the performance of our algorithm to that of other techniques. These comparisons show that the proposed algorithm can detect the vacancies in the parking spots while ignoring the false data and other distortions.

## I. INTRODUCTION

Vehicles are considered as one of the basic needs for people today. With increase in human populace, the vehicle populace has also expanded, causing heavy traffic. One of the essential reasons has been long vehicle blockages on main streets and particularly while finding the empty places or openings for stopping in numerous urban areas[1][2]. Indeed, even today, in many urban areas, the current parking areas are exceptionally conventional and gives inaccurate data with respect to the parking availability to the drivers. This issue calls for genuine considerations for making smart parking systems[3]. Existing parking spaces incorporate manual systems of parking where a parking officer helps the driver find an empty place for the vehicle. It turns out to be much more problematic for the officer to oversee if the parking zones and spaces increase[4]. Therefore, these problems can be solved by designing an intelligent, smart parking system. With so much time being wasted in heavy traffic and searching for parking spaces smart parking systems are necessary. Generally, two methods are used for the location of empty spaces: sensor based and picture based frameworks. Compared to sensor based parking systems, image based systems are more reliable, accurate and cost-effective[5][6]. Image-based parking systems can provide information to the drivers either using display boards at the entrance of every parking section or directly send a message to their mobile phones. A scalable solution would be to use the existing and widely-deployed video surveillance camera networks, which requires only the development of computer vision algorithms which would enable them to detect the vacant parking spots[7][8].

In this paper, a novel method for processing the image and removing distortions from the segmented image have been proposed. The method proposed shows better results than the existing methods and displayed high accuracy when tested in real time scenarios. In the conventional visual based segmentation procedures such as Gaussian processing, Edge based methods and Block Based Classification methods the results while accurate cannot be applied for all the cases and show false results while dealing with color specific cases[9][10]. It is observed that in most cases, detecting light colored cars in brightly illuminated settings has been problematic, but with the proposed algorithm the color of the vehicle and the illumination doesn't affect the accuracy of the result. The proposed method with the combined techniques of background illumination correction and false contour detection is robust enough to work with most of the cases dealt.

## II. IMAGE SEGMENTATION AND PRE-PROCESSING

### A. Image Processing

Pre-processing is widely used in operations with images at the lowest level of abstraction. These images are similar to the original data captured by the sensor, with the image represented as a matrix of image function values. The goal of pre-processing is to improve the image data that suppresses unwilling distortions or enhances some image features which are important for further processing. Geometric transformations of images also are among pre-processing methods due to the similarity in the techniques used.

Before processing the image, the non-uniform illumination of the image is corrected for better recognition of the objects in the image. The goal of illumination correction is to remove uneven illumination of the image caused by sensor defaults, non-uniform illumination of the scene, or orientation of the objects surface. Illumination correction is based on background subtraction. This method assumes that the image consists of a homogeneous background and comparatively small objects which are brighter or darker than the background.

### B. Segmentation and Labelling

Binary Image segmentation is an important part of image analysis to split the image into binary form. Segmenting an image to its binary form displays the prominent features of the image. Segmenting is usually carried out with the help of image histogram. The most common image property to threshold is pixel grey level: $M(x,y) = 0$ if $M(x,y) < Thr$ and $M(x,y) = 1$ if $M(x,y) \geq Thr$, where $Thr$ is the threshold. Using two thresholds, $Th1 < Th2$, a range of grey levels related to area1 1 can be defined: M(x,y) = 0 if M(x,y) < Th1 OR M(x,y) > Th2 and M(x,y) = 1 if Th1 ≤ M(x,y) ≤ Th2.

Connected components labeling, scans the image pixels for its connected components and groups the pixels with respect to its connectivity. Once all the groups have been determined, all the pixels are labeled according to the component it was assigned to. Extracting and labeling of disjoint and connected components in an image is critical to many autonomous image processing applications.

## III. METHODOLOGY

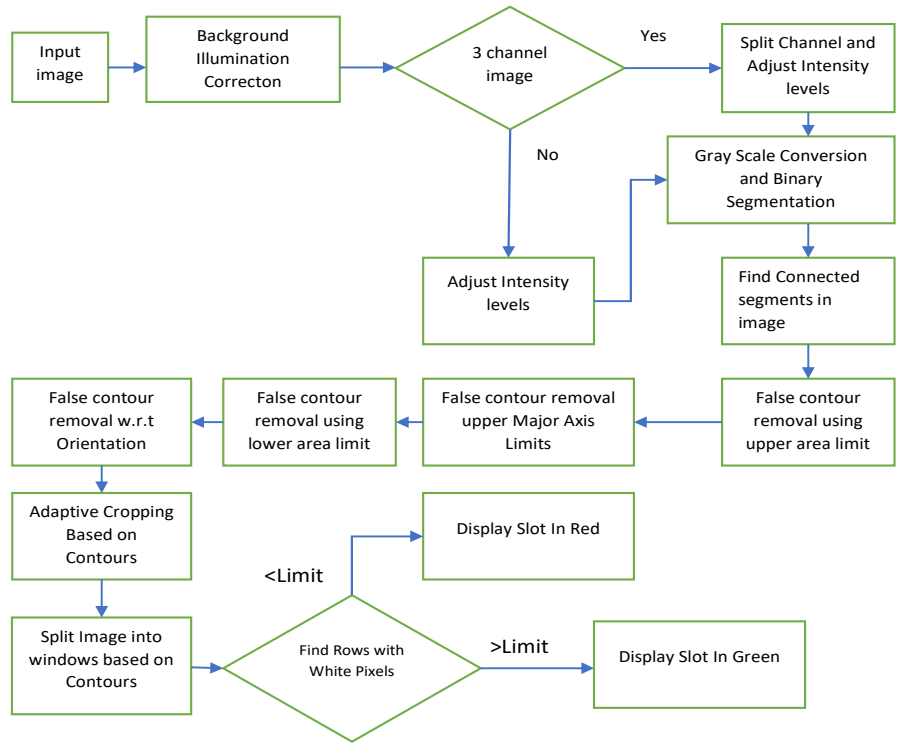

### A. Correcting Non-Uniform Illumination

Illumination correction is based on the background subtraction. The background illumination is estimated, and the background is subtracted from the original Image. After correcting the non-uniform illumination, the image is split into it's corresponding RGB channels. The intensity values of each channel are adjusted for optimal display of image characters and merged back into single image.

Figure 1: shows the background approximation of the input image. In the Figure 1, [0, 0] represents the origin, or upper-left corner of the image. The highest part of the approximation indicates that the highest pixel values of background and the lowest pixel values occur at the bottom of the image.

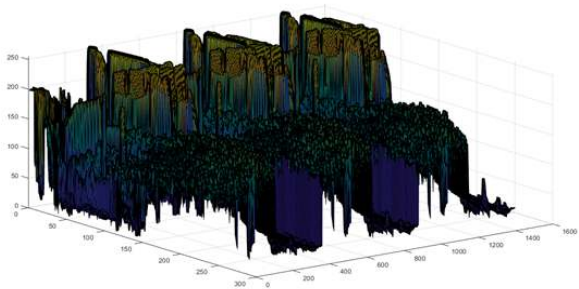

Figure 1: Background Approximation

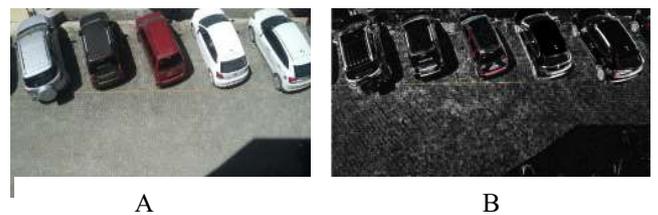

Figure 2A Input image, Figure 2B Background and Illumination Corrected Image

### B. Image Segmentation

Figure 2B Background and Illumination Corrected Image is converted into its respective gray image and then segmented into a binary image. The holes in the image are a set of background pixels that cannot be reached by filling in the background from the edge of the image.

The image is then scanned for the connected regions after which the thickening of the objects takes place by adding pixels to the exterior of objects. This results in the previously unconnected objects being 8-connected to a certain desired amount while preserving the Euler number. This process increases the prominent non-zero pixels in the image for easy recognition of the features.

A

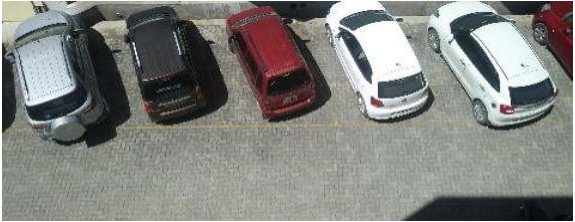

B

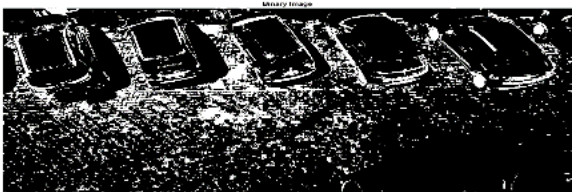

Figure 3A Input image, Figure 3B Binary Image

### C. Connected Components and Region Properties

The connected components in Figure 3B Binary Image are searched for using 8-connectivity. 8-connected pixels are neighbors to every pixel that touches one of their edges or corners. These pixels are connected horizontally, vertically, and diagonally. The connected components in the image are labeled accordingly in random order given that each label contains set of 8-connected pixels.

The regional properties of the labels such as Area, Major axis length and Pixel coordinates are calculated and stored in the form of a structure. Area refers to the total number of pixels in each region, Major axis length is the longest diameter connecting the edge points of a given region.

### D. Removing Undesired Components and Enhancing Prominent Features

The Figure 3B Binary Image has lot of unnecessary objects which come in the way of detecting an automobile efficiently. The region of interest and other regions differ prominently by three factors, which are a) area of the regions b) major axis length c) Orientation Angle. By introducing a threshold for area and major axis length, the undesired elements of the Figure 3B Binary Image can be removed. The connected regions below the threshold 3000, major axis length of 300 and above 1.7645e+03, and orientation between 0 to10 units are masked and assigned a false value. This enables only the connected components of the automobiles to be visible in the image which makes the classification easier.

Morphological transformation is used to enhance the features of the cars in the Figure 3B Binary Image. Two linear structuring elements and a diamond shaped structural element are created, which are an essential part of morphological dilation and erosion operations. A flat structuring element is a binary valued neighborhood, in which the true pixels are included in the morphological computation while excluding the false pixels. The center pixel of the structuring element, identifies the pixel in the image being processed.

The linear structuring elements have angles of 0 and 90 degrees, while the diamond shaped structural element has a radius of 1 unit. The linear structuring element is used for dilation and the Diamond shaped structural element is used for eroding the image.

A

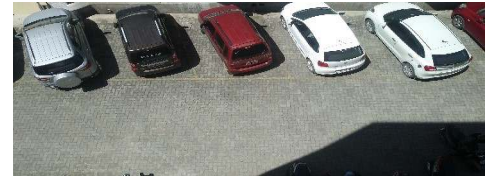

B

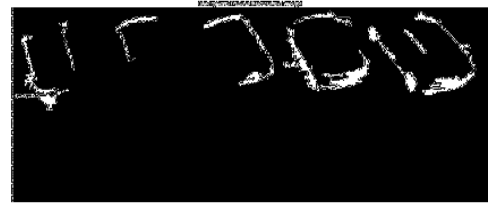

Figure 4A Input image, Figure 4B Binary Image After Noise Removed

### E. Adaptive Cropping

The Figure 4B Binary Image After Noise Removed, is split primarily into 3 regions vertically, each region is scanned for non-zero pixels and if the non-zero pixels are found to be less than the threshold of 10 pixels in the region, that region is cropped out of the Figure 4B Binary Image After Noise Removed. This process repeats in a loop each time reducing the area of the scanning window until a threshold of the window reaches 500 units and then terminates.

This process takes place if the Figure 4B Binary Image After Noise Removed, has at least one car present or else manual values are used to crop the image depending on the position of the parking lanes.

A 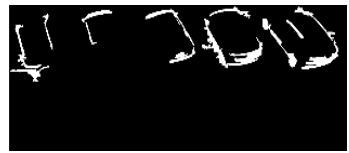  B 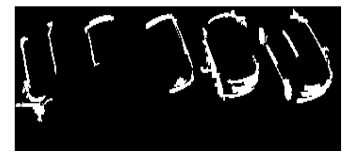

Figure 5A Figure 4B Binary Image After Noise Removed, Figure 5B Binary Image after Adaptive Cropping

## IV. EXPERIMENTAL RESULTS

The Figure 5B Binary Image after Adaptive Cropping is split into equal region horizontally depending on the number of the parking slots. The connected regions in Figure 5B Binary Image after Adaptive Cropping are detected by using a sliding window method. The window slides over each region and detects the number of non-zero pixels in the region. If the region has non-zero pixels over a threshold of 30 pixels, then the region as labelled as occupied or else the region is free for parking.

Figure 6　Result by proposed method

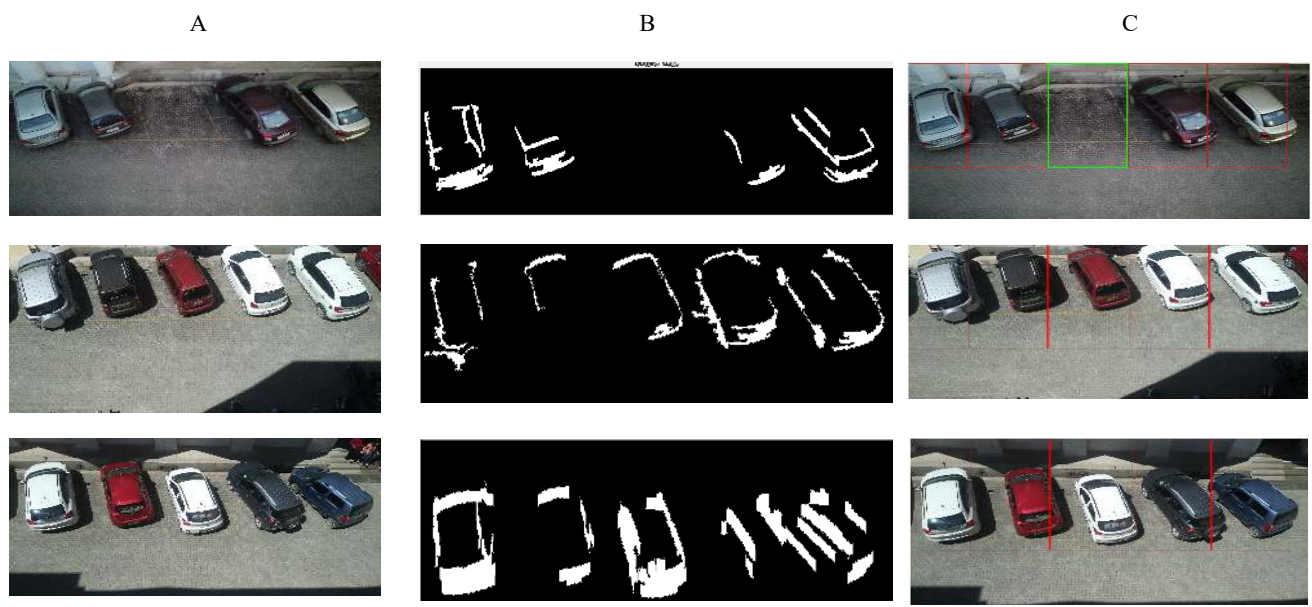

Figure 6A input Images, Figure 6B Binary Images with Car Contours, Figure 6C Results using the Proposed Method

Figure 6C shows the results obtained using images from different locations and time points. It can be observed that colors of the car nor the illumination of the image present any problem with the results. The detection of white cars, which has been problematic in most algorithms has been detected with accuracy in the proposed method.

Figure 7 Result Comparison Between Different Methods

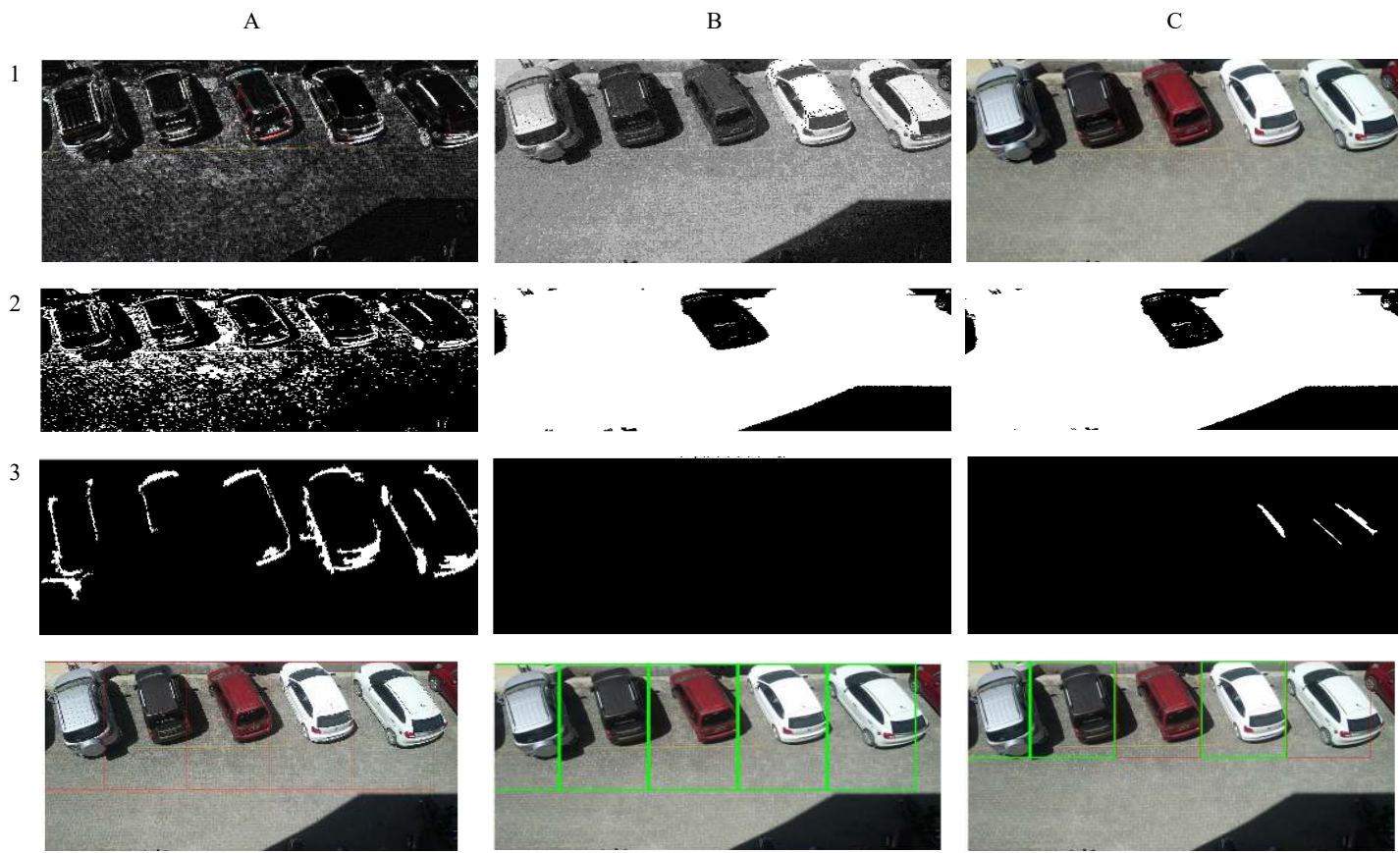

Figure 7,1 input Images, Figure 7,2 Binary Images, Figure 7,3 Binary Image After Noise Removal, Figure 7,4 Results obtained
Figure 7A Using Proposed Method, Figure 7B Using Gaussian Blur, Figure 7C Without Background Illumination Adjusted
Figure 7,4 displays the results in comparison of the proposed method with results obtained from gaussian blur and without using the pre-processing techniques. It can be observed that the proposed method gives accurate results.

Figure 8 Segmentation without proposed method

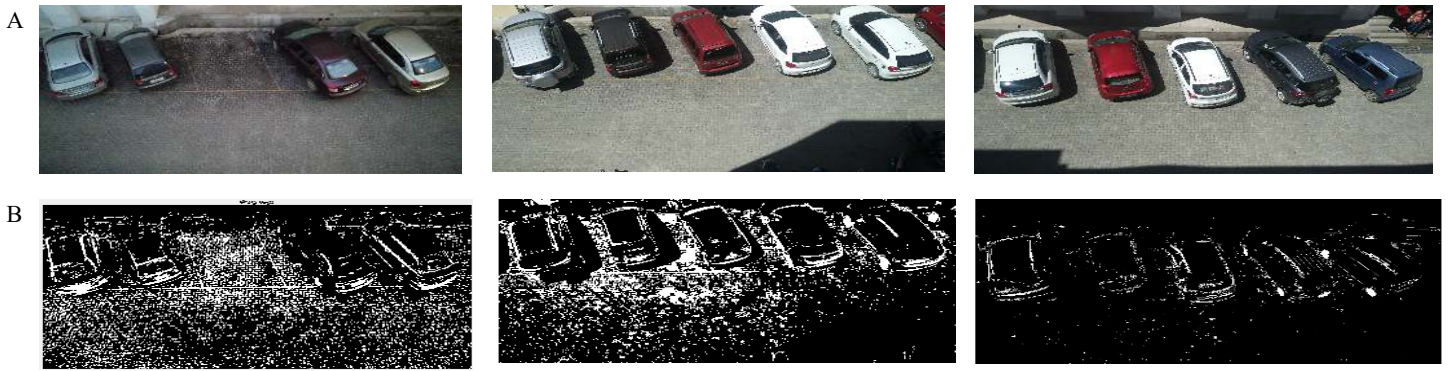

Figure 8A Input Image, Figure 8B Binary Image without using the proposed algorithm

It can be observed in Figure 8B Binary Image without using the proposed algorithm, that without using the proposed method to remove the false contours we get lot of noise.

To test the performance of our proposed algorithm, the accuracy of the system is measured with images taken at different time intervals. The performance is calculated by comparing the results of occupancy to the ground truth after every 5 sec. The performance of the proposed system is measured by the using the equation (1)

Accuracy percentage =

(Number of correct slots detected/ Total slots) *100

Table 1

| Vehicle appearance | Tests Performed | Correct Detections | False Detection | Accuracy |
|---|---|---|---|---|
| Clear | 42 | 40 | 2 | 99.5% |
| Occluded | 53 | 53 | 0 | 100% |

The accuracy of the proposed algorithm is found to be 100%, 99.5% accurate. The Table1 show less efficiency during clear sky, because of the bright illumination. Excess noise merged with the contours of the car, decreases the efficiency and the accuracy for detections. It is observed that the average performance is 99.57 % and is very high as compared with other parking lot detections applications. The accuracy of the proposed work also depends on the type of camera used for monitoring the parking lot.

Table 2

| | Edge Detection[10] | Twin ROI [12] | Block Based Classification [11] | Without Background illumination correction | Proposed method |
|---|---|---|---|---|---|
| Detection % | 97% | 95% | 98% | 60% | 99.57% |

From Table 2 we can conclude, the accuracy of our proposed system is better than the Gaussian Blur, Edge based detection technique[10], Block Based Classification technique[11] and Twin ROI methods used in existing parking and image segmentation techniques. In other methods, the efficiency goes down when the car and parking area is of the same color and the illumination is higher.

## CONCLUSION

Conventional algorithms to detect parking slots don't work with this particular area, due to the unclear, distorted parking lines and high illumination. This led to the implementation of a new method to detect the cars in the image based on the connected sets and contours after correcting the background illumination and removing false contours in the image. The proposed method shows promising results in such environments which give excessive noise when segmented, with areas of unclear boundaries and areas with low illuminations. The results indicated are accurate showing if the car is present or not in the given slot.

While the results are accurate, there are few changes which can improve the ease of use. The initial number of windows while calibrating the final results can be detected autonomously. The cropping ratio only takes into consideration of vertical cropping, implementing horizontal cropping along with the vertical cropping would be advantageous. The bounding boxes are not exactly aligned with the cars and are not angled. Finding the exact bounding fit for the cars and angling the visual boxes would enhance the appeal of the image.

## REFERNCES

[1] Mamta Bachania, Umair Mujtaba Qureshia, Faisal Karim Shaikha, "Performance Analysis of Proximity and Light Sensors for Smart Parking", The 7th International Conference on Ambient Systems, Networks and Technologies (ANT 2016)

[2] Ching-Fei Yang, You-Huei Ju, Chung-Ying Hsieh, Chia-Ying Lin, Meng-Hsun Tsai, Hui-Ling Chang, "iParking a real-time parking space monitoring and guiding system", Vehicular Communications 9 (2017) 301-305

[3] V. Stornelli, G. Ferri, M. Muttillo, L. Pantoli, A. Leoni, G. Barile, D. DâOnofrio, F.R. Parente and T. Gabriele," Wireless Smart Parking Sensor System for Vehicles Detection"

[4] Nazia Bibi, Muhammad Nadeem Majid, Hassan Dawood, and Ping guo," Automatic Parking Space Detection System", 2017 2nd International Conference on Multimedia and Image Processing

[5] Markus Heimberger a, Jonathan Horgan b, Ciarán Hughes b, John McDonald b, Senthil Yogamani, "Computer vision in automated parking systems:Design, implementation and challenges", Image and Vision Computing 68 (2017) 88-101

[6] Markus Heimbergera, Jonathan Horgan, Ciaran Hughesb , John McDonaldb , Senthil Yogamanib," Computer Vision in Automated Parking Systems: Design, Implementation and Challenges", Image and Vision Computing

[7] Piotr Dalka, Grzegorz Szwoch, and Andrzej Ciarkowski," Distributed Framework for Visual Event Detection in Parking Lot Area"

[8] Yucheng Li, Guohui Li, Xingai Xao, "Approach for Parking Spaces Based on ARM Emedded System"

[9] E.Tsomko, H.J.KIM, E.Izquierdo, "Linear Gaussian blur evolution for detection of blurry images"

[10] Ms.Sayanti Banerjee, Ms. Pallavi Choudekar, Prof .M.K.Muju, "Real Time Car Parking System Using Image Processing"

[11] Nazia Bibi, Muhammad Nadeem Majid, Hassan Dawood, and Ping guo "Automatic Parking Space Detection System"

[12] Najmi Hafizi, Bin Zabawi Sunardi, Kamarul Hawari Ghazali, Parking lot detection using image processing method, October 2013.